\documentclass[aps,prl,reprint,showpacs,groupedaddress]{revtex4-1}

\usepackage{graphicx}
\usepackage{amsmath}
\usepackage{amssymb}
\usepackage{color}
\usepackage[normalem]{ulem} 

\newcommand{\bremove}[1]{{\color[rgb]{0.6,0,0}\sout{#1}}} 

\renewcommand{\bremove}[1]{} 

\begin{document}

\title{Spin-echo dynamics of a heavy hole in a quantum dot}

\author{X. J. Wang, Stefano Chesi, W. A. Coish}

\affiliation{Department of Physics, McGill University, Montreal, Quebec H3A 2T8, Canada}

\date{\today}

\begin{abstract}
We develop a theory for the spin-echo dynamics of a heavy hole in a quantum dot, accounting for both hyperfine- and electric-field-induced fluctuations.  We show that a moderate applied magnetic field can drive this system to a motional-averaging regime, making the hyperfine interaction ineffective as a decoherence source. Furthermore, we show that decay of the spin-echo envelope is highly sensitive to the geometry.  In particular, we find a specific choice of initialization and $\pi$-pulse axes which can be used to study intrinsic hyperfine-induced hole-spin dynamics, even in systems with substantial electric-field-induced dephasing. These results point the way to designed hole-spin qubits as a robust and long-lived alternative to electron spins.
\end{abstract}


\maketitle

Electron spins in solid-state systems provide a versatile and potentially scalable platform for quantum information processing~\cite{Loss1998,Hanson2007}. This versatility often comes at the expense of complex environmental interactions, which can destroy quantum states through decoherence. Many theoretical and experimental studies have now established that the coherence times of electron spins in quantum dots \cite{Khaetskii2002,Merkulov2002,Petta2005,Bluhm2010}, bound to donor impurities \cite{Desousa2003,George2010}, and at defect centers \cite{Childress2006} are typically limited by the strong hyperfine interaction with surrounding nuclear spins~\cite{Hanson2007,Coish2009}. Heavy-hole spin states in III-V semiconductor quantum dots have emerged as a platform that could mitigate the negative effects of the hyperfine interaction. Due to the $p$-like nature of the valence band in III-V materials, the contact interaction vanishes for hole spins, leaving only the weaker anisotropic hyperfine coupling \cite{Fischer2008,Eble2009}. Moreover, the anisotropy of this interaction in two-dimensional systems should allow for substantially longer dephasing times in a magnetic field applied transverse to the quantum-dot growth direction \cite{Fischer2008,Coish2009}.

Recent experiments have measured hyperfine coupling constants for holes~\cite{Chek2011a,Chek2011b,Fallahi2010}, as well as spin-relaxation ($T_1$) \cite{Gerardot2008} and free-induction decay times, $T_2^*$, through indirect (frequency-domain)~\cite{Brunner2009} and direct (time-domain) studies~\cite{DeGreve2011}. Coherent optical control has now been demonstrated for hole spins in single~\cite{DeGreve2011,Godden2012} and double quantum dots~\cite{Greilich2011}. This technique has been used to implement a Hahn spin-echo sequence~\cite{DeGreve2011} giving an associated spin-echo decay time, $T_2\sim 1\,\mu\mathrm{s}$. The $T_2$ value reported in Ref.~\cite{DeGreve2011} has been attributed to device-dependent electric-field fluctuations, rather than the intrinsic hyperfine interaction. Motivated by these recent experiments, here we present a theoretical study of heavy-hole spin-echo dynamics with an emphasis on identifying the optimal conditions for extending coherence times. In particular, we show that dephasing due to electric-field fluctuations, as proposed in Ref.~\cite{DeGreve2011}, is dramatically suppressed in an alternate geometry considered here. Moreover, in contrast with the case of electron spins, we find that hole spins can enter a motional-averaging regime in a moderate magnetic field. In this regime, coherence is no longer limited by the hyperfine interaction, solidifying the potential for long-lived hole-spin qubits.

\begin{figure}
\includegraphics[width = 0.4\textwidth]{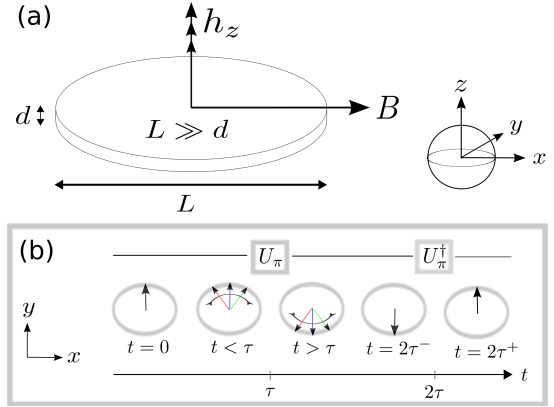}
\caption{(Color online) (a) Quantum-dot geometry, with nuclear field $h_z$ and magnetic field ${\bf B}=B \hat x$. For unstrained and flat quantum dots ({$d\ll L$}), $\gamma_H = g_\perp\mu_\mathrm{B}\simeq 0$ and $h_{x,y}\simeq 0$ \cite{Fischer2008,Fischer2010,Maier2012}. (b) Hahn echo sequence. Two $\pi$-rotations ($U_\pi$ and $U_\pi^\dag$, taken here about the $x$-axis) are applied at $t=\tau$ and $2\tau$, to refocus the HH spin.}
\label{fig:geometry}
\end{figure}

We consider a heavy-hole (HH) spin interacting with nuclear spins in a flat quantum dot with weak strain (see Fig. \ref{fig:geometry}). The HH spin is then described with the following Hamiltonian \footnote{In Eq. \eqref{eq:fullH}, we neglect terms $\sim h_{x,y}$, and nuclear quadrupole coupling. This is valid for a flat quantum dot with weak strain \cite{Fischer2008,Fischer2010,Maier2012,Sinitsyn2012}. The nuclear dipolar interaction may also influence the Hahn-echo decay of an electron spin on a timescale $\tau_{dd}\sim 10\,\mu\mathrm{s} \propto 1/\sqrt{A^{(e)}}$ \cite{Yao2006}. Since $|A|<|A^{e}|$, we expect $\tau_{dd}$ to be still longer for holes, beyond the times considered here.} (setting $\hbar=1$),
\begin{equation}
\label{eq:fullH}
H = H_{Z}+ h_z S_z,\quad H_{Z}=-\gamma_H B S_x - \displaystyle\sum\limits_{k} \gamma_{i_k} B I_k^x, 
\end{equation}
where $\mathbf{S}=\boldsymbol{\sigma}/2$ is a pseudospin-$1/2$ operator in the two-dimensional ($J^z=\pm 3/2$) HH subspace and $\mathbf{I}_k$ the nuclear spin at site $k$. $H_Z$ gives the hole- and nuclear-Zeeman interactions for an in-plane magnetic field ${\bf B}= B \hat{x}$ [see Fig. \ref{fig:geometry}(a)]. $\gamma_{i_k}$ is the gyromagnetic ratio of isotope $i_k$ at site $k$ having total spin $I_{i_k}$. The hole gyromagnetic ratio is $\gamma_H = g_\bot \mu_B$, with $g_\bot$ the in-plane $g$-factor for a dot with growth axis along [001] and $\mu_B$ the Bohr magneton. The hyperfine interaction \cite{Fischer2008,Coish2009} is expressed in terms of the Overhauser operator, $\mathbf{h} = \sum_{k} A_k \mathbf{I}_k$. The coupling constants, $A_k$, are given by $A_k = A^{i_k} v_0 |\psi(\mathbf{r}_k)|^2$, with $A^i$ the hyperfine constant for isotope $i$, $v_0$ the volume occupied by a single nuclear spin, and $\psi(\mathbf{r}_k)$ the HH envelope wavefunction. When the isotopes are distributed uniformly across the dot, we define the average $A = \sum_k A_k \simeq \sum_i \nu_i A^i$, with $\nu_i$ the isotopic abundance. In this case, and for a Gaussian envelope function in two dimensions, $A_k \simeq (A/N) e^{-k/N}$ \cite{Coish2004} with $N=10^4 - 10^6$ a typical number of nuclear spins within a quantum-dot Bohr radius. The ratio of $|A|$ to the strength of the hyperfine coupling of electrons, $|A^{(e)}|$, has been estimated theoretically~\cite{Fischer2008} in GaAs and confirmed experimentally \cite{Chek2011a,Fallahi2010} in InGaAs and InP/GaInP to be $|A/A^{(e)}| \sim 0.1$. For simplicity, we will evaluate numerical estimates with a single averaged value $|A^i| \simeq |A| \simeq 13 \, \mu \text{eV}$ \cite{Fischer2008,Coish2009} and $\nu_i,\gamma_i$ appropriate for $\rm In_{0.5}Ga_{0.5}As$. 

\paragraph{Spin echo.}
Under the action of Eq.~\eqref{eq:fullH}, spin dephasing results from fluctuations in $h_z$. Provided these fluctuations remain static on the timescale of decay of the hole spin, this source of decay can be removed via a Hahn echo [see Fig.~\ref{fig:geometry}(b)].
The process is better analyzed in the interaction picture with respect to $H_Z$, 
\begin{equation}\label{H_interaction_picture}
\tilde{H}(t) = \tilde{h}_z(t)\tilde{S}_z(t),
\end{equation}
where, for any $\mathcal{O}$, $\tilde{\mathcal{O}}(t)=e^{iH_Z t}\mathcal{O}e^{-iH_Z t}$. In particular, $\tilde{h}_z(t) = \sum_k A_k[I_k^z \cos{(\gamma_{i_k} B t)} - I_k^y\sin{(\gamma_{i_k} B t)}]$ and $\tilde{S}_z(t)=[S_z\cos (\gamma_H B t) - S_y\sin (\gamma_H B t)]$. The time-evolution operator after a time $2 \tau$ is then given by:
\begin{equation}\label{eq:U}
\tilde{U}(2\tau) = \mathcal{T} e^{-i\int_0^{2\tau} dt \tilde{H}_e(t)}.
\end{equation}
Here, $\mathcal{T}$ is the time-ordering operator and the modified echo Hamiltonian,
\begin{align}
\tilde{H}_e(t) &= \begin{cases}
\tilde{H}(t) &~~~~ 0\leq t < \tau,\\
\sigma_\alpha \tilde{H}(t) \sigma_\alpha &~~~~ \tau\leq t \leq 2\tau, 
\label{eq:HIecho}
\end{cases}
\end{align}
takes into account $\pi_\alpha$-pulses ($\pi$-rotations about $\alpha=x,y,z$). As seen in Eq.~(\ref{eq:HIecho}), $\pi_x$-pulses (but in general not $\pi_y,\pi_z$) have the beneficial effect of inverting the sign of the Hamiltonian, $\tilde{H}(t) \to -\tilde{H}(t)$, in the interval $\tau\leq t \leq 2\tau$. Provided $\tilde{H}(t)$ is approximately static over the interval $0<t<2\tau$, this will induce time-reversed dynamics for $\tau\leq t \leq 2\tau$, refocusing decay at the time $2\tau$. For this reason, unless otherwise specified, we will focus in the following discussion on a geometry with the magnetic field along $\hat{x}$ and $\pi_x$-pulses. We will contrast this analysis later with an alternate geometry relevant to recent experiments.

\begin{figure}
\includegraphics[width = 0.45\textwidth]{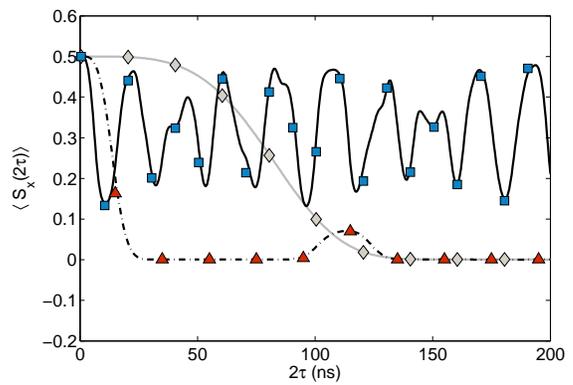}
\caption{\label{fig:exact}
(Color online) Exact analytical spin-echo decay with $B = 50\,{\rm mT}$ (solid gray), $B=2 \,{\rm T}$ (dash-dotted), and $B=10\,{\rm T}$ (solid black), corresponding to the regimes $ B \lesssim A/(\gamma_i\sqrt{N})$, $ B \simeq A/(\gamma_i\sqrt{N})$, and $ B \gtrsim A/(\gamma_i\sqrt{N})$, respectively. Markers show the approximate form, Eq.~(\ref{eq:magnus}). We have chosen $\left<S_x(0)\right> = 1/2$, $N = 10^4$, $\gamma_H = g_\bot\mu_B = 0$, and $\gamma_i$ from Table 1 of Ref.~\cite{Coish2009}.}
\end{figure}

\paragraph{Vanishing $g_\bot$.}
We first consider the limit $\gamma_H = g_\bot\mu_\mathrm{B} = 0$ in Eq.~(\ref{eq:fullH}). The dynamics we find in this limit will be a good description whenever $\gamma_H< \gamma_i$, corresponding to $g_\perp<10^{-3}$ ($g_\perp<5\times 10^{-3}$ has been reported in 2D wells \cite{Korn2010}). This limit considerably simplifies the theoretical analysis and allows for an exact solution: $H$ becomes block diagonal in the eigenbasis of $S_z$ and, in each block, the eigenstates are obtained after rotating $I_k^z$-eigenstates by an angle $\theta_k = \pm \arctan(2 \gamma_{i_k} B/A_k)$ about $\hat{y}$. Representative results of the exact evolution of $\langle S_x(2\tau) \rangle$ are shown in Fig.~\ref{fig:exact}. The spin-echo signal has a remarkable dependence on the magnetic field: there is a clear transition from a low-field regime, where the decay time decreases with increasing $B$, to a high-field regime, where there is no decay, only modulations of the echo envelope. 

To give physical insight, we have developed an analytical approximation scheme based on the Magnus expansion. The Magnus expansion is an average-Hamiltonian theory typically applied to periodic and rapidly oscillating systems \cite{Maricq1982}. This scheme is suggested by the oscillating terms in Eq.~(\ref{H_interaction_picture}), and will allow us to analyze the more general problem with $\gamma_H \neq 0$. In the Magnus expansion, we assume the evolution operator, Eq.~(\ref{eq:U}), can be written as
$\tilde{U}(2\tau) = e^{-iH_M(2\tau)} =e^{-i\sum_{i=0}^\infty H^{(i)}(2\tau)}$.  The $i^\mathrm{th}$-order term, $H^{(i)}(t)$, is found using standard methods \cite{Maricq1982}. Each higher-order term in the Magnus expansion contains one additional integral over time.  Oscillating terms are therefore suppressed by a factor of order $\| \tilde{H} \|/\omega$, with $\omega$ the typical oscillation frequency. The leading-order term is $H^{(0)}(t) = \overline{H(t)}t$, where $\overline{H(t)}$ is simply the average of $\tilde H_e(t)$ over an interval $t$. The spin components $S_\alpha$ ($\alpha = x,y,z$) are then given by:
\begin{equation}\label{eq:Liouvillian}
\big\langle S_{\alpha}(2\tau)\big\rangle
= \big\langle \tilde{U}^{\dagger}(2\tau) \tilde{S}_{\alpha}(2\tau) \tilde{U}(2\tau) \big\rangle 
= \big\langle e^{iL_M(2\tau)} \tilde{S}_{\alpha}(2\tau) \big\rangle , 
\end{equation}
 where $L_M(t)$ is defined by $L_M(2\tau)\mathcal{O} = \big[H_M(2\tau), \mathcal{O}\big]$
and $\langle\mathcal{O}\rangle ={\rm Tr}\{\mathcal{O} \rho\}$. The initial state $\rho = \rho_S \otimes \rho_I$ is assumed to describe a product of the hole-spin ($\rho_S$) and nuclear-spin ($\rho_I$) density matrices, where the nuclear spins are in an infinite-temperature thermal state. For $N\gg 1$ uncorrelated nuclear spins, the central-limit theorem gives nearly Gaussian fluctuations, resulting in
\begin{equation}\label{eq:momexp}
\big\langle e^{iL_M(2\tau)} \tilde{S}_{\alpha} \big\rangle \simeq \bigg\langle \text{exp}\Big\{-\frac{1}{2}\big\langle L_M^2(2\tau)\big\rangle_I \Big\} \tilde{S}_{\alpha}\bigg\rangle_S, 
\end{equation}
where we define $\langle L_M^2(t)\rangle_I\mathcal{O}_S = {\rm Tr}_I\{(L_M^2(t)\mathcal{O}_S) \rho_I \}$ and $\langle\mathcal{O}\rangle_S = {\rm Tr}_S\{\mathcal{O} \rho_S\}$. 

At high $B$, rapid oscillations in $\tilde{H}(t)$ allow us to keep only the leading term: $L_M(2\tau)\mathcal{O}_S \simeq L^{(0)}(2\tau)\mathcal{O}_S = [H^{(0)}(2\tau),\mathcal{O}_S]$. Setting $\gamma_H=0$, as appropriate for Fig.~\ref{fig:exact}, and with the help of $\sum^\prime_k A_k^2\simeq \nu_i A^2/(2N)$ (where the prime restricts the sum to nuclei of isotopic species $i$), we obtain:
\begin{equation}\label{eq:magnus}
 \frac{\langle S_x(2\tau) \rangle} {\langle S_x(0) \rangle}\simeq
\exp{\left[ -\sum_i \frac{4\nu_i A^2 I_i(I_i+1)}{3 N (\gamma_i B)^2} \sin^4{\left(\frac{\gamma_i B \tau}{2}\right)}\right]}.
\end{equation}
As seen in Fig.~\ref{fig:exact}, Eq.~(\ref{eq:magnus}) (markers) reproduces the exact dynamics very well. The precise conditions for the validity of the Magnus expansion will be given below.

\begin{table}
\centering
\begin{tabular}{|c|c|c|c|}
\hline
& ~$\gamma_H \gg \gamma_i$~ & ~$\gamma_H \ll \gamma_i < \frac{A}{B\sqrt{N}}$~ & ~$\gamma_H \ll \frac{A}{B\sqrt{N}} < \gamma_i$~ \\ \hline\hline
$\omega$~ & ~$\sim B\gamma_H$~ & ~$\sim B\gamma_i$~ & ~$\sim B\gamma_i$ \\
$\delta \omega_{\text{rms}}$~ & ~$ \sim A/\sqrt{N}$ ~&~ $\sim A/N$ ~&~ $\sim A/N$ \\
$\tau_{\text{max}}$ ~&~ $\sim \omega/\delta \omega^2_{\text{rms}}$ ~&~ $\sim \frac{1}{A}(\omega/\delta \omega_{\text{rms}})^3$ ~&~ $\sim \omega/\delta \omega^2_{\text{rms}}$  \\ \hline
\end{tabular}
\caption{\label{tab:MagnusValidity} The Magnus expansion will generally reproduce the correct dynamics for $\delta\omega_\mathrm{rms}/\omega <1$ and $\tau<\tau_\mathrm{max}$, with $\omega$, $\delta\omega_\mathrm{rms}$, and $\tau_\mathrm{max}$ given above in three regimes.}
\end{table}

The simple form of Eq.~(\ref{eq:magnus}) enables us to understand why the behavior of $\langle S_x(2\tau) \rangle$ changes as $B$ is increased. For $ B \ll A/(\gamma_i\sqrt{N})$ (gray solid line in Fig.~\ref{fig:exact}), a short-time expansion of Eq.~(\ref{eq:magnus}) gives $\langle S_x(2\tau) \rangle \simeq \langle S_x(0) \rangle \left(1-(\tau /\tau_o)^4\right) \simeq \langle S_x(0) \rangle e^{-(\tau /\tau_o)^4}$, with
\begin{equation}\label{tau_o}
\tau_o \simeq \frac{1}{\sqrt{B}}\left[\sum_i \frac{\nu_i ( \gamma_i A)^2}{4N} \frac{I_i(I_i+1)}{3}\right]^{-1/4}.
\end{equation}
Surprisingly, when $B$ is increased, $\tau_o$ \emph{decreases}.  This behavior is opposite to the situation for electron spins, in which the echo decay time \emph{increases} for increasing $B$ \cite{Cywinski2009}. This decrease is due to rapid fluctuations in $h_z$ from nuclear spins precessing at frequencies $\sim \gamma_i B$.  The Hahn echo can no longer refocus these dynamical fluctuations at finite $B$, although Eq. \eqref{eq:magnus} does predict partial recurrences (dash-dotted line in Fig. \ref{fig:exact}) due to the finite number of discrete precession frequencies $\sim\gamma_i B$.   

\begin{figure}
\begin{center}
\includegraphics[width = 0.45\textwidth]{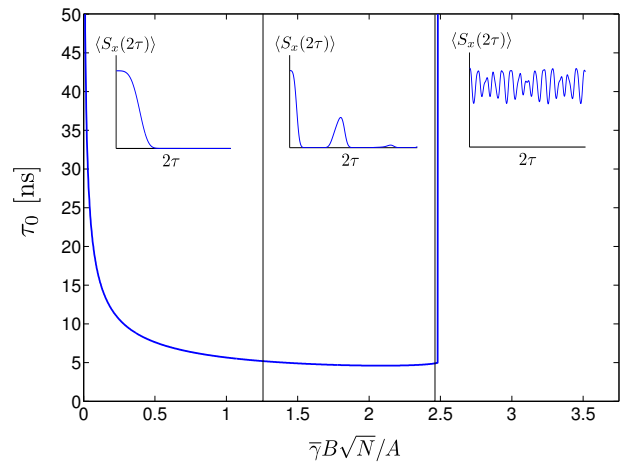}
\caption{\label{fig:decaytime}
(Color online) Decay time, $\tau_o$, vs. $B$. Here, $\overline{\gamma}=\sum_i \nu_i \gamma_i$ and parameters are as given in the caption of Fig.~\ref{fig:exact}. Insets: typical $\langle S_x(2\tau) \rangle$ in each of the three regions: $\overline{\gamma} B \sqrt{N}/ A\ll 1$, $\overline{\gamma} B \sqrt{N}/ A\sim 1$, and $\overline{\gamma} B \sqrt{N}/ A\gg 1$. }
\end{center}
\end{figure}

In contrast, at large magnetic field, $B \gtrsim A/(\gamma_i \sqrt{N})$, the system enters a motional-averaging regime in which the decay of $\langle S_x(2\tau) \rangle$ is bounded by $\sim (A/B\gamma_i\sqrt{N})^2$, giving rise to beating (black solid curve in Fig. \ref{fig:exact}). This beating has the same physical origin as electron-spin-echo envelope modulation (ESEEM) \cite{Rowan1965}, although the extreme anisotropy of the hole hyperfine interaction allows uniquely for its complete suppression.  Fig.~\ref{fig:decaytime} shows the $1/e$ decay time, $\tau_o$, as $B$ is increased, leading to a discontinuity when $\gamma_i B \sqrt{N}/A\gtrsim 1$, at which point $\left<S_x(2\tau)\right>$ always remains close to its initial value [see Eq.~(\ref{eq:magnus})].

\paragraph{Finite $g_\bot$.}
Although there are definite advantages to making flat unstrained dots leading to $g_\bot\simeq 0$ and $h_{x,y}\simeq 0$, current experiments are performed on hole systems with finite (albeit small) $g_\bot$ \cite{Marie1999,DeGreve2011}. For this general case, with $\gamma_H = g_\bot\mu_B \neq 0$, we have no closed-form exact solution for the dynamics, but our analysis can still be applied for a certain range of $\tau, B$. 

We neglect subleading oscillating terms in the Magnus expansion when $\|(H^{(0)})^2 \| \gg \|H^{(0)}H^{(2)} \|, \|\left(H^{(1)}\right)^2\|$. More specifically, if the relevant fast oscillation frequency is $\omega\sim\gamma_{i} B$, each precessing nuclear spin experiences a typical hyperfine field $\delta \omega_{\rm rms} \sim A/N$ from the hole. Otherwise, if the fast frequency is $\omega\sim\gamma_H B$, the hyperfine field acting on the precessing hole is of order $\delta \omega_{\rm rms} \sim A/\sqrt{N}$, averaging over the nuclear configurations. As a consequence, the parameter $\delta \omega_{\rm rms}/\omega < 1$ controls the expansion with $\delta\omega_\mathrm{rms}$ and $\omega$ given in Table \ref{tab:MagnusValidity} for each regime.  In addition to bounded oscillating terms, the Magnus expansion generates terms that grow with $\tau$.  These terms approach $\sim 1$ at $\tau\sim\tau_\mathrm{max}$, beyond which a finite-order Magnus expansion may fail.  Nevertheless, the Magnus expansion will provide an accurate description whenever $\delta\omega_\mathrm{rms}/\omega<1$ and for $\tau \lesssim \tau_\mathrm{max}$.  Estimates of $\tau_\mathrm{max}$ are given in Table \ref{tab:MagnusValidity}.  The sufficient conditions presented here may be overly conservative in specific cases. For example, the parameters of the $B=50$ mT curve of Fig.~\ref{fig:exact} give a short $\tau_{\rm max}<0.1$ ns, while the Magnus expansion is clearly valid up to a much longer time scale.  This is, however, a fortuitous example; we find that in analogous calculations of free-induction decay, the bounds are tight.

\begin{figure}
\includegraphics[width = 0.45\textwidth]{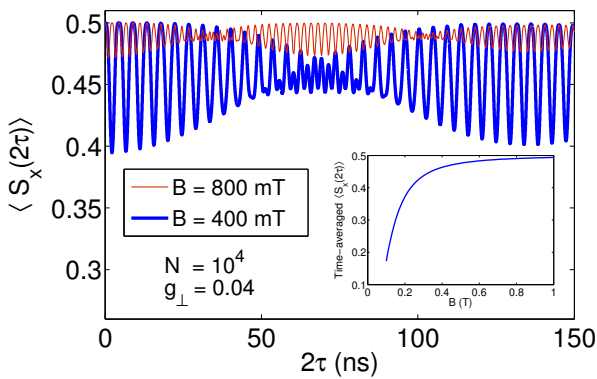}
\caption{(Color online) Main panel: spin-echo envelope from a leading-order Magnus expansion when $\gamma_H\gg\gamma_i$ in the motional-averaging regime, $\gamma_H B > A/\sqrt{N}$. Inset: Time-averaged $\left<S_x(2\tau)\right>$. We have taken $g_\bot=0.04$ \cite{Marie1999}, $\left<S_x(0)\right> = 1/2$, and $N=10^4$. These results are valid at least for $2\tau \lesssim 2\tau_\mathrm{max}\simeq 70\,\mathrm{ns}\,(140\,\mathrm{ns})$ for $B=400\,\mathrm{mT}\,(B=800\,\mathrm{mT})$.}
\label{fig:motional_narrowing}
\end{figure}

In practice, the value $g_\bot = 0.04$ measured in \cite{Marie1999} suggests that $\gamma_H \gg \gamma_{i}$ in many current experiments. For $g_\bot=0.04$, the condition $B > A/(\gamma_H \sqrt{N})$ is already satisfied above a rather small value, $B \gtrsim 100$~mT for $N\sim 10^4$. In Fig.~\ref{fig:motional_narrowing} we plot representative curves in this motional-averaging regime, displaying the same features discussed for $g_\bot =0$. Additionally, fast oscillations at the hole Zeeman frequency, $\gamma_H B$, induce beating in the echo envelope function, $\langle S_x(2\tau) \rangle$, which is not present for $g_\bot = 0$. 

\paragraph{Decay anisotropy.}
While the results discussed so far are specific to $\pi_x$-pulses, other schemes are possible. Due to the extreme anisotropy of the hole-spin hyperfine coupling, the spin-echo decay is also highly anisotropic, depending on both the initialization and $\pi$-pulse axes. If the hole spin is initialized along a generic in-plane direction $\hat n=n_x\hat{x}+n_y\hat{y}$, and $\pi$-rotations are performed about that same axis, we find that $\langle {\bf S}(2\tau) \cdot \hat{n}\rangle / \langle {\bf S}(0) \cdot \hat{n}\rangle$ is independent of $\hat n$ when $g_\bot=0$. This result is to be expected since any in-plane component of the hole spin experiences the same effective field, $\tilde{h}_z(t)$, along the $z$-axis. On the other hand, for $g_\bot \neq 0$, rotational symmetry about the $z$-axis is broken, resulting in a strong in-plane anisotropy. For the parameters of Fig.~\ref{fig:motional_narrowing}, but with initialization along $\hat{y}$ and $\pi_x$-pulses, we obtain that $\langle S_y(2\tau) \rangle$ is dominated by the hole Larmor precession about the $x$-axis and approaches the simple sinusoidal function $\langle S_y(2\tau) \rangle \simeq \langle S_y(0) \rangle \cos(2\gamma_H B \tau)$ in the motional-averaging regime, $\gamma_H B> A/\sqrt{N}$. 

Additional dephasing mechanisms other than the nuclear bath can also have a strong influence on the precession about $\hat{x}$, introducing other sources of anisotropy. In particular, the decay of $\langle S_y(2\tau) \rangle$ was measured in \cite{DeGreve2011} with $\pi_z$-pulses used for the Hahn echo. The resulting decay was found to be approximately exponential, $\langle S_y(2\tau) \rangle \simeq \langle S_y(0) \rangle e^{-2\tau /T_2}$, with a $B$-independent $T_2 \sim 1\,\mu$s. This behavior was attributed to spectral diffusion induced by electric-field noise, which we model here by setting $\gamma_H B \to \gamma_H B+\delta\omega(t)$ in Eq.~(\ref{eq:fullH}). The observed exponential decay is consistent with Gaussian white noise \cite{Klauder1962} $\langle \delta \omega(t)\delta \omega(t') \rangle_{\delta \omega} = \frac{2}{T_2} \delta(t-t')$ (and $\langle \delta \omega(t) \rangle_{\delta \omega}=0$), where $\langle \ldots \rangle_{\delta \omega}$ indicates averaging with respect to realizations of $\delta\omega(t)$. We have included this additional dephasing mechanism in the evaluation of Eq.~(\ref{eq:momexp}) for the $\pi_x$-pulse echo sequence examined previously and obtained a power-law decay at $\tau \gg T_2$:
\begin{equation}\label{g_noise_result}
\frac{\langle S_x(2\tau) \rangle }{\langle S_x(0) \rangle }\simeq \langle \exp\left[\frac{\langle h_z^2 \rangle_I }{2}\sum_{\alpha=y,z} f_\alpha^2(t) \right] \rangle_{\delta \omega} \simeq \frac{1}{1+ \tau/\tau_D},
\end{equation}
where $ f_y(\tau) =\int_0^{2\tau} \sin\phi(t) {\rm sgn}(\tau-t)dt $, $f_z(\tau) =\int_0^{2\tau} \cos\phi(t) dt$, $\phi(t)=\gamma_H B t + \int_0^{t} \delta\omega(t') dt'$, and 
\begin{equation}\label{tau_D}
\tau_D=\frac{1 + (\gamma_H B T_2)^2}{2\langle h_z^2 \rangle_I T_2}.
\end{equation}
This decay time scale is exceedingly long ($\tau_D \simeq 20$ s) for the experimental value $\gamma_H B \simeq 2\times 10^{11}\,{\rm s}^{-1}$ and using $\langle h_z^2 \rangle_I \sim 10^{15}\,{\rm s}^{-2}$, which demonstrates the negligible effect of spectral diffusion on the previous discussion (e.g., Figs.~\ref{fig:exact}, \ref{fig:decaytime}, and \ref{fig:motional_narrowing}). For simplicity, we have derived Eqs.~(\ref{g_noise_result}) and (\ref{tau_D}) with static nuclear-field fluctuations $\langle h_z^2 \rangle_I $. This corresponds to a worst-case scenario for the present model. At the high magnetic field of Ref.~\cite{DeGreve2011} ($B \sim 8$ T), motional averaging would likely inhibit decay even further. 

\paragraph{Conclusion.}
We have calculated the spin-echo dynamics of a single heavy-hole spin in a flat unstrained quantum dot. The relevant dynamics are highly anisotropic in the spin components and $\pi$-rotation axes. When $\gamma_H \ll \gamma_{i}$, we predict an initial decrease of the coherence time with increasing $B$, followed by a complete refocusing of the HH-spin signal and motional averaging when $B\gtrsim B_c$ ($B_c\sim A/\gamma_i\sqrt{N}\simeq 3\,\mathrm{T}$ for $N=10^4$). The motional-averaging regime is also realized when $\gamma_H \gg \gamma_{i}$, relevant to current experiments. In this regime, decay due to the hyperfine coupling can only occur for $\tau\gtrsim \tau_\mathrm{max}\propto B$, and can therefore be completely suppressed. We have further shown that device-dependent electric-field noise becomes negligible for a specific geometry, allowing for a measurement of the limiting intrinsic decoherence due to nuclear spins.  We expect the systematic approximation scheme introduced here to find wide applicability to a number of other challenging spin dynamics problems associated with nitrogen vacancy centers, donor impurities, and electrons in quantum dots.

We acknowledge financial support from NSERC, CIFAR, FQRNT, and INTRIQ.

\bibliography{holebib}

\end{document}